# Teaching Software Process Modeling

Marco Kuhrmann  
Technische Universität München  
Faculty of Informatics  
Garching, Germany  
kuhrmann@in.tum.com

Daniel Méndez Fernández  
Technische Universität München  
Faculty of Informatics  
Garching, Germany  
mendezfe@in.tum.com

Jürgen Münch  
University of Helsinki  
Department of Computer Science  
Helsinki, Finnland  
juergen.muench@cs.helsinki.fi

*Abstract*—Most university curricula consider software processes to be on the fringes of software engineering (SE). Students are told there exists a plethora of software processes ranging from RUP over V-shaped processes to agile methods. Furthermore, the usual students' programming tasks are of a size that either one student or a small group of students can manage the work. Comprehensive processes being essential for large companies in terms of reflecting the organization structure, coordinating teams, or interfaces to business processes such as contracting or sales, are complex and hard to teach in a lecture, and, therefore, often out of scope. We experienced tutorials on using Java or C#, or on developing applications for the iPhone to gather more attention by students, simply speaking, as these are more fun for them. So, why should students spend their time in software processes? From our experiences and the discussions with a variety of industrial partners, we learned that students often face trouble when taking their first "real" jobs, even if the company is organized in a lean or agile shape. Therefore, we propose to include software processes more explicitly into the SE curricula. We designed and implemented a course at Master's level in which students learn why software processes are necessary, and how they can be analyzed, designed, implemented, and continuously improved. In this paper, we present our course's structure, its goals, and corresponding teaching methods. We evaluate the course and further discuss our experiences so that lecturers and researchers can directly use our lessons learned in their own curricula.

## I. INTRODUCTION

Software process models are the glue that holds organizations, projects, and people together. Therefore, the development, the maintenance, and the improvement of a software process model constitute challenging tasks that each requires well-trained and experienced process engineers. Although being an essential part of software engineering, software processes are, if at all, only marginally included in a typical software engineering (SE) curriculum. Typically, such curricula contain lectures and labs that consider the SE basic principles, requirements engineering, architecture, programming and programming languages, and some in-depth courses, which are usually specific to the universities' focus, e.g., automotive software engineering, business information systems, or robotics.

The university curricula, as they are defined, provide the basic computer science knowledge including a rudimentary SE toolbox. They lack, however, the opportunity to experience "real world" problems, e.g., programs of considerable size in large-scale development projects with moaning customers and further soft facts that realistically shape typical problems and risks. Hence, students are often left unprepared for todays industrial project environments. This is problematic as most of the software projects fail not because of technical issues but because of an insufficient ability to understand project organization and management, which is basically reflected by software processes. Studies show that *"with few exceptions, the reasons that large-scale development programs have failed have not been technical [...]. As the cancellation of two large and critical efforts demonstrates, these systems have almost always failed because of program-management problems"* [1].

Those topics that cause the majority of economic damage to industry[1] are not taught at university in a way students can experience the effects of missing or poorly implemented software processes. The reason for this shortcoming is that the aforementioned aspects in SE need to be tailored in order to fit into the lectures' schedule. This is additionally enforced by strict time schedules in the curricula, e.g., arising from the Bologna process[2]. On the other hand, it can be seen as fruitful with respect to learning if students can experience themselves the effects of the SE principles or techniques/methods they are taught. This might help to better understand their practical relevance. Programs, for instance, can be developed and tested by the students themselves. Errors that were made during the development can be easily experienced. Also, for lecturers it is easier to monitor, correct, and rate those kinds of tasks. Software processes, on the other hand, are more complicated to teach. Since the "execution engine" of software processes is an organizational context involving human beings, software processes cannot be taught the same way like programming languages.

While todays SE curricula mostly address the system layer, software processes are situated on the organization and project layer in which completely different topics matter. However, since students will work usually after graduation in the context of large projects that are coordinated by significantly complex processes (whether they are explicitly defined or not), they need to have a fundamental understanding about process models and process management (e.g., how processes can be used to coordinate teams), as well as of challenges and risks that are associated with software processes and their deployment or improvement.

---

[1]IDC's worldwide IT investment volume forecast in 2010 estimated a growth to 1,735 Trillion $ in 2013—at the same time the Standish Group periodically names about 25% of the projects to be a failure.

[2]The Bologna Process – Towards the European Higher Education Area, http://ec.europa.eu/education/higher-education/bologna_en.htm

*A. Problem Statement*

Although we have insights into todays industrial project environments with all its challenging and problematic facets, we still miss to effectively prepare the students for those environments. Students need to be prepared for such environments with a basic understanding about the relevance and the goals of software process modeling and management. The design of a course that gives students access to this understanding and the evaluation of such a course's success remains a challenging task, mainly because of soft facts given in the envisioned context that cannot be generalized and taught in classical courses structured with lectures and exercises only.

Yet missing are concepts and lessons learned for the design and the evaluation of a course on software process modeling including the analysis of processes, their implementation, and their improvement.

*B. Objective*

We aim at defining a concept for a course that pays attention to the skills a process engineer must have in today's industrial environments. Therefore, we aim at defining general goals and requirements for a course that applies to the needs of teaching software processes. We design and implement the structure and the content of such a course in a universities curriculum, and validate our approach against the requirements.

*C. Contribution*

We contribute a concept and its evaluation for a course on software process modeling going beyond a classical teaching format with lectures and exercises. To this end, we define the requirements for a course to tackle the problems stated above. We contribute the design of the course via a blueprint including most relevant topics and a proposed schedule.

The concept has been fully implemented and evaluated at the Technische Universität München in the winter term 2011/2012. The evaluation includes the evaluation against the requirements, the formal evaluations according to the faculties standards, and an informal feedback-based evaluation performed by the students. In addition, parts of the concepts have been implemented in a process definition and management course (based on [2]) at the University of Helsinki (winter 2012/2013). Experiences of this implementation are integrated in the discussion section. Finally, we conclude with a discussion of our results.

*D. Outline*

The remainder of this paper is organized as follows: In Sect. II, we discuss the related work. Section III defines the requirements a course on software process modeling should meet. In Sect. IV, we describe the course design, introduce a blueprint, map the requirements, and show an example implementation. Section V presents an evaluation, which is based on a formal evaluation by the faculty and an informal one that is based on the students' feedback. We conclude the paper in Sect. VI by critically discussing the passed courses and drawing a roadmap for future work.

## II. RELATED WORK

Different concepts have been proposed to teach software process aspects. Besides traditional classroom teaching, we distinguish between approaches based on practical exercises, experiments, games, and simulation. The Personal Software Process (PSP) [3] is the most prominent teaching approach that focuses on practical exercises. The underlying idea is to apply process principles at the level of single developers. Studies have shown that applying such processes at individual level can lead to significant performance improvements [4]. In contrast to the approach presented here, the PSP exercises all deal with relatively small and local processes.

Empirical approaches focus on teaching by conducting experiments, typically as part of a regular course. The students take the role of experimental subjects to experience the effects of selected processes or techniques themselves. Typical objectives of such experiments consider comparisons of different quality assurance processes (e.g., [5]), Global Software Engineering (GSD, e.g., [6]–[8]), or Software Engineering in general (e.g., [9]–[12]). In contrast to the approach presented here, the experimental treatments are usually engineering level processes and not process modeling or process management activities. Only few experiments exist that focus on process management aspects (e.g., [13], [14]).

Approaches for teaching lean and agile practices are often based on educational games [15]. Lean production processes, for instance, are often demonstrated with the means of a game to impart knowledge about lean principles. In contrast to the approach presented in this paper, this kind of teaching typically aims at a better understanding of a specific philosophy rather then at better understanding the challenges of defining and managing large software processes.

Finally, simulation is sometimes used to support teaching in the area of software processes [16]. Here, students can make local decisions and see their global effects. Simulation is also suited for playing "what if games" and, thus, help to better understand processes. In contrast to the our approach, simulation have a more limited scope that consists of understanding process dynamics.

All these different approaches have their specific strengths and could be seen, at least partially, as valuable additions to the approach presented in this paper.

## III. REQUIREMENTS FOR A PROCESS MODELING COURSE

Software engineering comprises many tasks—technical ones as well as organizational and management tasks. A number of roles in projects are responsible to perform those tasks and to produce artifacts. Furthermore, several stakeholders are also included in a project, e.g., top management, IT services, interdisciplinary users. Because of the number of participating people and the projects' objectives, coordination and communication are, besides the classical engineering disciplines, essential to meet the project goals. Software process models are a means to describe how the communication is structured, how collaboration is established, which artifacts have to be produced, and how to coordinate teams to operate (manage and control) a software project in general.

### A. Basic Goals

The basic goal of the course is to effectively teach the basic concepts in software process modeling over their whole life cylcle, i.e. the analysis of processes, the conceptualization/design of processes, their tool-supported implementation, and, finally, their deployment and evaluation.

### B. Requirements

Based on the basic goals, we derived the following requirements for a course on software process modeling. The requirements basically originate from experiences gathered in a number of process improvement and coaching projects and trainings. In the following, we provide brief explanations why we consider those requirements to be important.

*REQ 1: Students need to understand the importance of process models*

Students will work in organizations and projects where they have to deal with processes with all their facets. Hence, they need an understanding why things work the way they do. Even in the case in which students work in an environment that is free of any (documented) process, students need to have knowledge about project organization, collaboration pattern, and opportunities to operate projects successfully, for example, considering progress control.

*REQ 2: Students need to understand the integration of process models into an organizational environment*

We believe that students need to learn how organizations are structured in general. Such knowledge includes interfaces between a project and their embedding into an organization (e.g., project internal processes and their relation to business processes), knowledge about process groups (e.g., the role of the quality management), and process infrastructures and their requirements (e.g., organizational and cultural prerequisites to apply agile methods like Scrum or to implement a process model using SPEM/EPF).

*REQ 3: Students need to experience realistic process models in terms of differences in size and complexity related to small, medium, and large organizations and projects*

We often observe that educational programs are focused on agile development practices in small teams. However, it is still factually important to teach that software processes are a means to organize software development, especially when applying agile methods. Still, depending on the organizations' culture or the project size, different process models are of interest and need to be taught. For instance, a globally acting enterprise usually implements a process model of considerable size and complexity to reflect the structure, in general, and certain project situations, in particular. Such process models may contain hundreds of work products and activities, or dozens of roles, which, if summarized, we have experienced to constitute thousands of process elements to be maintained (see, e.g., the V-Modell XT).

*REQ 4: Students need differentiated knowledge about different modeling approaches*

Since there exists a plethora of modeling approaches that can be used to design and implement a process model, students need to understand those approaches, their basic philosophies, concepts, and the basic idea behind a particular approach. Each modeling approach has opportunities and limitations, which can become "show stoppers", because they might be incompatible to the organization's philosophy. A wrong design decision can lead to a valid process model, which might, however, not be implementable; for example, if the chosen process modeling tool does not support certain association types. Also, weak designs may cause increased effort in terms of maintainability. Therefore, the different approaches need to be effectively taught in conjunction with their consequences for, e.g., developing/realizing a process or identifying and correcting errors.

*REQ 5: Students need to learn the most relevant existing process models (such as national or international standards)*

Organizations often use or have to some extent adhere to existing process models. Students need to understand the basic concepts of such models, their suitability for specific contexts, their underlying principles, and their customization needs. The selection of the most relevant process models depends on several factors (such as the country or a specific domain that might be addressed by the overall computer science curriculum). Examples might be international standards (such as ISO 12207:2008 [17]), national standards (such as the V-Model XT in Germany), or domain-specific standards (such as IEC 26262 [18] for the automotive domain or IEC 62304 [19] for medical devices).

## IV. COURSE DESIGN

At the university, we face the problem that the usual teaching format consists of a weekly lecture and an exercise (90 minutes each), where the exercise is usually done the week after the lecture. In consequence, theoretical parts are taught in a theater style, where the lecturer "acts" in front of the students. The week after, the topics of the lecture are repeated and supported with some exercises the students can work on. This leads to a work slot of effectively 90 minutes (at most) per week in which possible examples have to fit in. In terms of software process modeling, this is a ridiculous time span compared to real SPI projects, which comprehend an effort between 50 and more than 200 man days [20].

Trainings in an industry environment are differently structured. Such trainings are organized in a workshop style with typical durations of 3 to 5 days. In such trainings, theoretical and practical parts are interwoven. Lab-sessions of several hours can be included, which gives more freedom to work on examples. Of course, those workshops are still far away from "real" SPI projects, but proved to be more efficient than the dispersed university teaching style.

The basic course design, which we presented in [14], was developed according to the demand to change the way software processes are taught. From our experience, which we made in university teaching as well as in professional trainings in an industry context, we re-designed our lecture to meet the requirements from Sect. III.

In this section, we briefly introduce a software process life cycle model on which the course design is based and the resulting course design. We show how the requirements are met and present an example implementation.

### A. The Software Process Life Cycle

When teaching software process modeling to students, a systematic framework is required. Several of such software process life cycles (SPLC) have been proposed, e.g., an 8-step approach that describes the lifecycle of a descriptive process model [2]. These life cycles mainly differ in the span of the life cycle (number of covered phases) and the specific focus (prescriptive or descriptive modeling or process improvement). All these life cycle models follow a typical process, i.e. models can be specified, built, implemented, analyzed, used, assessed, evolved, or rejected.

TABLE I
PHASES IN THE SOFTWARE PROCESS LIFE CYCLE MODEL.

| Phase | Description |
|---|---|
| Analysis | In the analysis phase the actual process is analyzed. The process goals and the process requirements have to be determined. If available (e.g., when evolving a given process model), prescribed process models have to be analyzed. Techniques to be applied in this phase are for instance: interviews, workshops, or audits. |
| Conception | In the conception phase descriptive modeling techniques [2] are applied to create process prototypes. The overall goal of this phase is to define the target process model including, e.g., the selection of modeling techniques/adaptation options, creating drafts of the process to be, and so on. All those activities are done without paying much attention to concrete technical implementations. |
| Realization | In the realization phase the concrete realization approach is defined (including concrete modeling techniques that may be supported by a technical infrastructure, i.e. EPF [21]). Furthermore, the requirements and designs—made during the conception phase—are refined in order to be implemented in the selected (technical) environment. From the management perspective, the realization and the quality assurance plans are created on a level that allows for concrete implementation tasks, which are also performed in this phase. Depending on the kind and the complexity of the process model under consideration, the conception and the realization phases can be done in parallel. |
| Deployment | In the deployment phase pilot projects, trainings and concrete deployment strategies are concretized [22]. This includes the roll-out of a certain process in an organization or a specific change of an existing process that is already conducted in an organization. We also subsume in this phase an evaluation of the process by using different means such as assessment, audit, or measurement-based improvement. |

We use an aggregated and simplified life cycle for the course (see Table I), which consists of the following phases: analysis, conception, realization, and deployment. Since this life cycle is iterative, it supports for continuous improvement of software process models.

### B. Blueprint

Our course design[3] is based on both, our experience in university teaching as well as industry workshops. We adopted the workshop-based approach in the university teaching to give students more space to gain practical experience, and on the other hand, to give researchers the opportunity to conduct research [14]. The course concept (Fig. 1) uses the SPLC model from Table I as underlying conceptual framework to align all relevant topics into the course's structure. For the practical parts students go through the SPLC when analyzing, designing, and realizing a software process model. Because of the time constraints of a semester, practicing the deployment is not possible.

*1) Theoretical Parts (light gray):* The course consists of three lecture phases. In the first phase (about three weeks), the fundamentals and the basic knowledge is imparted. In the second phase of about eight weeks, the lecturer prepares the frame for the spotlight topics that cover areas of specialization. The students work independently on the spotlight topics; they prepare presentations and small essays to summarize their outcomes. Also, during this phase the practical parts are prepared. The third phase (about two weeks) is the evaluation phase. Outcomes from the practical parts are evaluated according to scientific methods. The topics as well as the proposed examples covered by the theoretical part can be taken Table II.

*2) Practical Parts (dark gray):* At the time when the second phase of the theoretical parts starts, the classical exercises, which can be used at the very beginning of the course over two sessions, are replaced by practical trainings (labs). Therefore, a project assignment in which the project objectives are contained needs to be prepared. The choice of the project should be opportunistic keeping, however, in mind that a realistic complexity should be inherent (an example is given in Sect. V-C). The project objectives need to be aligned with the course's contents (cf. Table II) w.r.t. the agenda of the overall course and topics that can be worked on in a self-contained manner, e.g., work packages that can be handled in one session. The chosen example should support a continuous and seamless work among the different sessions. Another important aspect is that outcomes of the practical parts need to be prepared for evaluation and, thus, need to be defined in order to be measurable. Figure 1 also shows the assignment of the SPLC phases to the second lecture phase and the workshops. In terms of the SPLC model, the workshops cover the analysis, the conception, and the realization as tasks that can be done in a workshop slot of 90 or 180 minutes.

---

[3]This teaching format was awarded with the "Ernst Otto Fischer Teaching Award" (2012) by the Faculty of Informatics, Technische Universität München (http://portal.mytum.de/studium-und-lehre/lehrpreise/ernst_otto_fischer_lehrpreis.html).

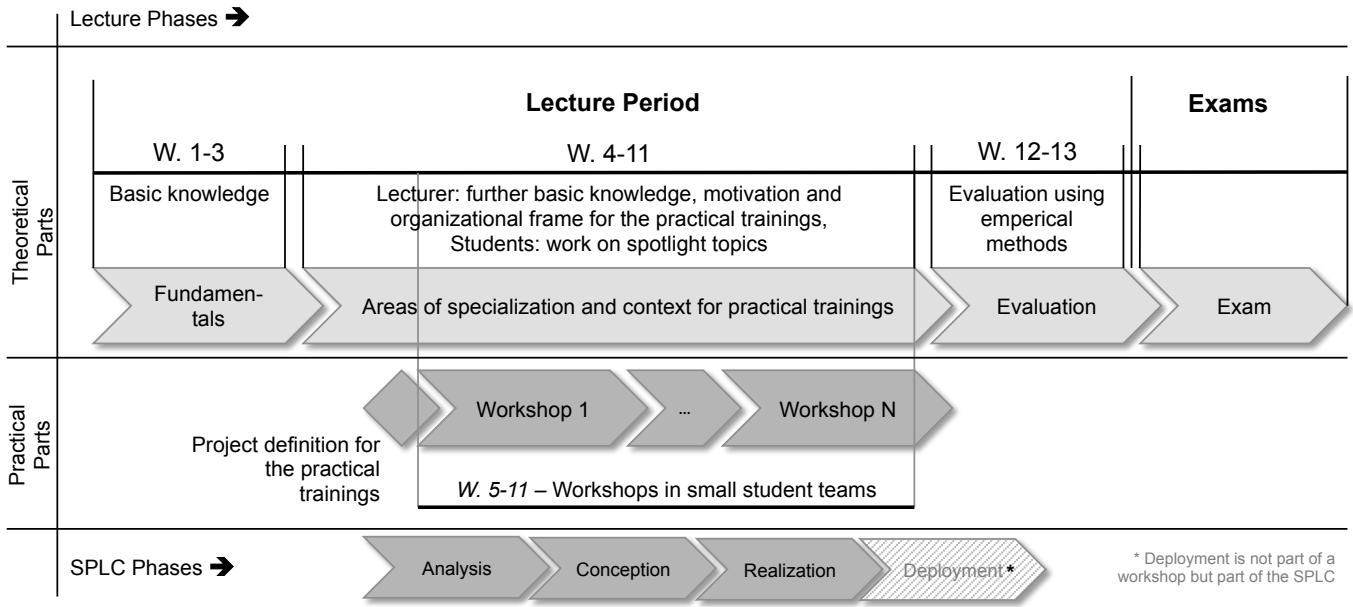

Fig. 1. Course blueprint showing the theoretical (light gray), the practical (dark dark) parts, the mapping to the SPLC phases, and a planning pattern.

*3) Scientific Parts:* We also introduced a pattern for conducting research in the context of this lecture format [14] including experimentation and survey research.

### C. Requirements Mapping

In Sect. III, we named the requirements we consider important to run a course on software process modeling. Furthermore, Fig. 1 shows the SPLC-based blueprint that gives the frame for such a lecture. In consequence, a number of topics needs to be addressed by the lecture that cover general information on software processes, the importance of software processes, their relation to organizations and other kinds of processes, as well as knowledge w.r.t. modeling approaches, concrete software process models, systematic approaches to analyze, design, realize, and deploy software processes, and supporting methods/techniques in order to do *software process engineering*. Therefore, we mapped the requirements, the blueprint, and the necessary topics using Table II. The table shows the lecture phases (fundamentals, areas of specialization, evaluation), names the performers (lecturers doing the theoretical parts, students working on spotlight topics, and the practical parts), and gives a mapping to the requirements. To support easy reproduction, the table also provides references to embody the key contents, such as concrete process models, selected methods, and tools.

### D. Example Implementation

The blueprint was implemented for the first time in the second run of the lecture[4] "Software Engineering Processes"

[4]Lecture "Software Engineering Processes", winter term 2011/2012, master's level, http://www4.in.tum.de/lehre/vorlesungen/vgmse/ws1112/index.shtml; material German and partially English, available on request

in the winter term 2011/2012 (the first lecture was given in the classic way). This lecture gives an introduction to the domain of process engineering and process management [2], [32], [37], [38]. Table II shows the topics addressed in the lecture aligned to the software process life cycle and the (common) tasks needed to be performed to analyze, conceptualize, design, implement, publish, and assess a software process. This course was implemented at the master's level. In order to provide a high-quality class, to foster interaction, and due to the experimental character, the group size was restricted to 15 students. The whole lecture was organized to be held in a block of 180 minutes per week to create the necessary space for practical trainings and experiments.

TABLE III
SCHEDULE FOR THE LECTURE "SOFTWARE ENGINEERING PROCESSES".

| Week | Topics |
| --- | --- |
| 1–3 | Fundamentals |
| 4–7 | Software processes and infrastructures, SPLC: Analysis, SPLC: Conception |
| 8–9 | Software process metamodels, SPLC: Realization, workshops: analysis and conception (90 minutes each) |
| 10–12 | SPLC: Deployment, SPLC: Management and continuous improvement, workshop: realization (180 minutes) |
| 13 | Evaluation |

A 13-week schedule for the pilot implementation is shown in Table III (in relation to Table II). The table boils down the blueprint (Fig. 1) and shows, what are the particular contents, and what is the duration a phase.

TABLE II
MAPPING BETWEEN LECTURE CONTENTS AND REQUIREMENTS (PERFORMERS: T = LECTURER (THEORY), S = STUDENT, P = PRACTICAL LAB)

| Performer | Topic in the lecture | REQ1 | REQ2 | REQ3 | REQ4 | REQ5 |
|---|---|---|---|---|---|---|
| **Fundamentals** | | | | | | |
| T | Motivation, need for software process models | ✗ | | | | |
| T | Software process terminology and philosophies | | | | ✗ | ✗ |
| T | Software processes and organizations (i.e. organization pattern) | ✗ | ✗ | | | |
| T | Business processes, software processes, and project management at a glance | ✗ | ✗ | | | |
| *Software processes and infrastructure* | | | | | | |
| T | Basic software process models (waterfall, V-shaped, iterative approaches) | ✗ | ✗ | | | ✗ |
| T/S | Agile methods (Scrum [23], XP [24], Crystal, FDD, Kanban, MSF Agile) | | | ✗ | | ✗ |
| T/S | Rich processes (RUP [25], V-Modell XT [26], Prince2, Hermes) | | | ✗ | | ✗ |
| T/S | Other approaches: (Situational) Method Engineering [27] | | | | ✗ | ✗ |
| T | Other processes: maturity models, e.g., CMMI [28], SPICE [29], ISO 9000 [30] | ✗ | | ✗ | | |
| T | Other processes: ITIL [31] and other processes for operation | | | ✗ | | ✗ |
| T | Tool-support for software process users: process enactment in projects | | | | ✗ | ✗ |
| T | Tool-support for software process authors: overview authoring tools | | | | ✗ | |
| **Areas of specialization** | | | | | | |
| T | The "Software Process Life Cycle" (SPLC, overview) | | | | ✗ | |
| *SPLC: Analysis* | | | | | | |
| T | Motivation and overview | ✗ | ✗ | | | |
| T/P | Stakeholder analyses, role models (context: organization pattern) | ✗ | ✗ | ✗ | | |
| T/P | Process and workflow analyses | | | ✗ | | |
| T/P | Artifact analyses | | | ✗ | | |
| T | Process model analysis (cf. descriptive modeling [2]) | | | ✗ | | |
| *SPLC: Conception* | | | | | | |
| T | Design strategies (overview) | | | | ✗ | |
| T | Conceptual modeling approaches (e.g., templates structures, etc.) | | | | ✗ | |
| T/P | Mapping of analyzed processes to process entities (e.g., milestones, roles, tasks) | | | | ✗ | |
| T | Design of process tailoring (overview over different approaches) | | ✗ | | ✗ | |
| T/P | Design of the process documentation | | | | ✗ | |
| T | Creating software processes by using software process lines [32] | | | ✗ | ✗ | |
| – | *Software process metamodels – see below* | | | | ✗ | |
| *SPLC: Realization* | | | | | | |
| T | General implementation strategies | ✗ | ✗ | | ✗ | |
| T | Iterative process realization approach | | | ✗ | ✗ | |
| T | Creating feedback loops and dealing with change | | | ✗ | ✗ | |
| *SPLC: Deployment* | | | | | | |
| T | Deployment strategies (big bang, incremental, pilot projects [22]) | ✗ | ✗ | ✗ | | |
| T | Planning of coaching and training | | ✗ | ✗ | | |
| T | Planning and set-up tools and tool infrastructures | | ✗ | ✗ | | |
| T | Experiences regarding "deployment traps" | ✗ | ✗ | ✗ | | |
| *SPLC: Management and continuous improvement* | | | | | | |
| T | Process management in the context of continuous improvement (overview) | ✗ | ✗ | ✗ | | |
| T | Introduction to continuous improvement (i.e. PDCA/Deming cycle [33]) | ✗ | | ✗ | | ✗ |
| T | Assessment/Audit/Certification for individuals, projects, and companies | | | ✗ | | ✗ |
| T/S | Selected models in detail (CMMI + Scampi [34], SPICE [29], Six Sigma) | | | ✗ | | ✗ |
| T | Planning and managing an improvement program | | | ✗ | | |
| *Software process metamodels and tools* | | | | | | |
| S | ISO 24744 (metamodel only [35]) | | | ✗ | | ✗ |
| S | SPEM (metamodel and tools, i.e. EPF, incl. hands on [36]) | | | ✗ | | ✗ |
| S | V-Modell XT (metamodel and tools, incl. hands on [26]) | | | ✗ | | ✗ |
| **Evaluation** | | | | | | |
| T | Overview empiric research (e.g., questionnaires, surveys – goals, how to's) | ✗ | | | ✗ | |
| P | Planning and performing an assessment | ✗ | | | ✗ | |

TABLE IV
FORMAL EVALUATION (COMPARISON WINTER SEMESTER 2010/2011 AND 2011/2012, TUM)

| Criterion | Winter semester 2010/2011 | Winter semester 2011/2012 | Result |
|---|---|---|---|
| Number of answered questionnaires | 6 | 8 | – |
| *Common criteria (1 = very high, 5 = very low)* | | | |
| Complexity | 3.00 (old questionnaire: "level") | 2.38 | +0.62 ↑ |
| Volume | 2.83 (old: one question) | 2.12 | +0.71 ↑ |
| Speed | | 2.75 | +0.08 → |
| Appropriateness of effort compared to ECTS points | n.a. | 3.00 (fair) | n.a. |
| *Overall rating (1 = very good, 5 = very bad)* | | | |
| Lecture | 1.25 | 1.5 | -0.25 ↘ |
| Exercise | 2.17 | 1.33 | +0.84 ↑ |
| Relation to practice | 2.0 | 1.62 | +0.38 ↗ |

## V. EVALUATION

In the following, we describe the evaluation of our course. Since we took into account the requirements stated in Sect. III, we omit the inference of formal criteria for their evaluation. This evaluation is inherently covered by the mapping shown in Table II. Instead, we rely on a formal evaluation predefined by our faculty and an informal evaluation directly performed anonymously by the students.

### A. Formal Evaluation

The faculty performs a formal evaluation, usually in the last third of the semester. Since we gave the lecture twice (first run in the classic fashion, second run in the new format), we can compare both runs. Due to updated questionnaires, the evaluations are not directly comparable. However, the basic information can still be extracted. The questionnaires are very comprehensive (8-10 pages) and contain questions regarding the lecture itself, the exercises, certain fine-grained aspects, such as evaluation of the lecturer, reasons for having missed some lectures, and so on. Regarding our requirements, we are especially interested in the following questions:

1) Did we manage to give a lecture in the new format, which stays on the same high level than before?
2) Did we get the students better motivated by the new style of the exercises (practical parts/labs, more interaction and so on)?
3) Were we able to sensitize the students about the practical relevance of process modeling?

To answer those questions, we analyzed the formal faculty questionnaires for appropriate criteria. Table IV shows the selected criteria.

Since not all students were present when making the evaluation, we have only few answered questionnaires (6 out of 9 from the first run, 8 of 14 form the second run) and, thus, those are not subject to quantitative analyses. The questionnaires, however, give answers to our questions. The selected common criteria show that the level of the lecture not only stays high (3 is the average in the scale), but also increases. Although changing the style of running this lecture, all common criteria show a (slight) improvement.

To answer the other questions, we identified three questions in the questionnaire that give an overall rating w.r.t. the lecture itself, the exercises, and the relation to practice. In the overall rating the lecture itself stays on a very high level while, at the same time, the rating of the exercises increases significantly. Our last question, whether we could better sensitize the students for the practical relevance of process modeling, also shows that the new format has to be considered successful.

Summarized, the ratings show that, in the second run, the lecture stays at a high level, but the quality of the exercises increases significantly whereby we have to consider that the students see the lecture also as more demanding. By introducing the new format that is focused on workshops and creative work rather than on classic exercises, students got closer to practice and, thus, could gather experiences.

### B. Informal Evaluation

In addition to the formal evaluation, we also performed an informal evaluation twice: the first evaluation after the first third of the lecture and a second evaluation at the end of the lecture. The evaluation consists of a *one-minute-paper*-like feedback. Each student should briefly fill out a small sheet of paper that covers the following aspects:

1) (up to 5) points that are positive
2) (up to 5) points that are negative
3) (up to 5) points that I still wanted to say (informal)

The structure of the class, the selected topics, the combination of theory and practice, the way of continuously evaluating the work and finding the final grades were rated positively. Especially the the practical projects and the team work in the workshops was highlighted. On the other hand, the students mentioned the tough schedule and the not always optimal tailoring of the tasks for the practical sessions. Also, the students wished to have more feedback loops w.r.t. presentation techniques and self-reflection.

Since we informed the students about the "experimental" character of this special course in advance, the students did not complain, but welcomed the opportunity to give the feedback to improve their own class.

## C. Experiences Made in Munich

We implemented the course twice (Munich: in 2010 in the classic format and in 2011 in the new format) and compared it to an equivalent course at the University of Helsinki (classic format with experience-oriented exercises). Since we implemented the course twice, we can compare both runs as well as draw some conclusions from our experiences.

Our experiences are two-fold: on the one hand, we can judge the teaching format itself, which we already did in [14]; on the other hand, we especially can compare the lecture runs according to, e.g., the level or the content. Fundamentally, our experiences are quite well reproduced by the faculty's evaluation, which is summed up in Table IV.

In addition to those ratings, we want to highlight the following experiences. In the following, we summarize our experiences.

*1) Real World Example:* We experienced that the choice of a real world example rather than a synthetical toy example has proved to be successful. For this class, we have been provided with the software process by the "Special Interest Group Software Processes" of the German Computer Society. The objectives of the workshops were to analyze this given process, to define several improvements, and to implement the design into two given software process frameworks (see also Table III). Our acceptance criteria (complete design documentation and process implementation) were achieved. Also, the process owner was satisfied giving us the feedback that he did not expect the student groups to create "such a comprehensive solution in this little time."

*2) Decisions & Consequences:* Another goal—"let students experience the consequences of their decisions"—was also achieved. While implementing the process in the workshop, we could observe a certain learning curve. For instance, one team had a complete design, but selected an inappropriate modeling concept in the Eclipse Process Framework [21]. The shortcomings became obvious when they tried to connect content packages and the delivery process, which was not supported the way the group thought it would be. The group had to refactor the implementation, which was an annoying and time-consuming task, both increasing their awareness on the consequences of certain design decisions.

*3) Forming a Team:* Another, not explicitly defined, goal was also achieved. Because of the team work, the considerable share of independent work and interaction in the class we formed working team. We observed the students improving their communication and collaboration skills. Much of the work was done due to intrinsic motivation of the teams.

*4) Exams:* We compared the final grades of both lectures and recognized significantly better grades in the second run. During the exams, the students could not only answer all (theoretical) knowledge-related questions, but also all knowledge transfer and application-related questions. The students usually referred to the practical examples and were able to transfer and apply their experiences to new situations.

## D. Experiences Made in Helsinki

In the course at the University of Helsinki, different types of practical assignments were used. Among more traditional exercises aiming at knowledge transfer, assignments such as process programming, process elicitation, and exploratory tasks were given to the students.

*1) Application of Modeling Paradigms:* Process programming was one of the topics of the practical assignments at the University of Helsinki. The students were asked to model different processes by using a formal notation. The students were able to better understand the level of detail of the required information (such as product or control flow information) that is necessary to formalize software processes. In addition, the students were asked to create process models for creative processes as well as for processes that can be fully automated. This supported, on the one hand, a better understanding of the typical characteristics of creative software processes and their differences to processes that are better suited for automation. On the other hand, students could experience that different types of processes require different modeling paradigms (such as a constraint oriented or an imperative paradigm).

*2) Utilization of Elicitation Techniques:* Another topic of the practical assignments at the University of Helsinki was process elicitation techniques. The students got interview transcripts from practitioners describing their activities in real projects. The students were asked to extract relevant process modeling concepts such as roles, artifacts, and responsibilities. This improved the understanding of the challenges when extracting process knowledge from real projects via interviews. The students experienced how difficult it is to transform informal information or tacit knowledge into process models. The students could also see how difficult it is for individuals to formulate their behavior in a rule-oriented manner.

*3) Suitability of Process Models for Different Contexts:* The course at the University of Helsinki included several exploratory assignments. The students were asked, for instance, about empirical evidence on the effects of specific processes in a certain environment (e.g., what kind of evidence exists for model-based testing w.r.t. its impact on reliability of the tested software in the automotive domain). These kind of exploratory tasks helped the students to see how many aspects are involved in process modeling and management. They also understood that there is not always only one solution or one best solution but that the choice of the most suited solution it depends usually on the context. The course used a conversation-based style for the exercises so that different students could present their solutions. The solutions were discussed afterwards. For some students it was surprising to see that typically several solutions are acceptable.

## VI. CONCLUSION & FUTURE WORK

In the following, we give a brief summary of conclusions and the implications we see in our course concepts, before concluding with an outline of future work.

We discussed the problem of missing to effectively prepare students for industrial life. Based on feedback from our industrial partners and our own experiences, we identified the missing education in software processes (and modeling) as a major shortcoming. Based on experience exchanges between the University of Helsinki and the Technische Universität München, we analyzed existing curricula, critically discussed our existing lectures on this topic, and derived key requirements courses on software process modeling should satisfy. We checked our courses against the proposed requirements and decided to design a new course concept having a new format to cope with the problems stated above.

The new concept allows for teaching the basic concepts of software process modeling with realistic scenarios at a realistic level of complexity supporting, for example, learning curves while establishing a differentiated knowledge about different modeling approaches. We evaluated our new lecture against our requirements by conducting a formal evaluation that follows our faculties standards and by conducting an informal evaluation in which the students directly rated the course. The direct comparison of our new teaching format with the classical format applied in previous terms showed, for example, a significant improvement of the exercises' quality and the relation of the contents to practice.

We consider this kind of lecture format to be successful for not only the field of software processes, but also for other software engineering fields. For the winter term 2012/2013 at the Technische Universität München, we are already conducting a lecture on agile project management methods following the same format[5]. Since we are confident of the suitability of the introduced concepts for teaching topics with a generally practical application field, it is further planned to replicate the format at courses with a completely different topic, e.g., in the area of empirical software engineering. Our course thus directly impacts on our own teaching curricula implemented in two different universities.

The contributed course concepts, the provided exemplary content map, and the experiences provided in this paper allows other lectures to optimize their own courses in the field of software process modeling. Furthermore, it also supports the establishment and optimization of courses on other topics that demand to apply practical examples, to experience real world problems, and to learn how to deal with those problems in a differentiated manner.

The presented course is a first experimental implementation step towards optimizing our courses on software process modeling. For this reason, we are working on extensions of our concepts in response to feedback from academia and industry, and on the transfer of the concept to other topics such as agile project management. Furthermore, we are planning a family of observational studies to investigate the effects this course has when being applied in other contexts or to a larger number of students. We thus cordially invite the reader to join the continuous improvement of the concepts and the future empirical investigation of the effects this teaching format has.

---

[5]Lecture "Agile Project Mangement & Software Development", winter term 2012/2013, master's level, http://www4.in.tum.de/lehre/vorlesungen/vgmse/ws1213/index.shtml; material available on request

## REFERENCES

[1] Humphrey, W. S., Konrad, M. D., Over, J. W., and Peterson, W. C., "Future directions in process improvement," *Crosstalk – The Journal of Defense Software Engineering*, vol. 20, no. 2, 2007.
[2] Münch, J., Armbrust, O., Soto, M., and Kowalczyk, M., *Software Process Definition and Management*. Springer, 2012.
[3] W. S. Humphrey, *PSP(SM) - A Self-Improvement Process for Software Engineers*. Addison-Wesley, 2005.
[4] Rombach, D., Münch, J., Ocampo, A., Humphrey, W. S., and Burton, D., "Teaching disciplined software development," *International Journal of Systems and Software*, vol. 81, no. 5, 2008.
[5] Kamsties, E. and Lott, C., "An empirical evaluation of three defect-detection techniques," in *5th European Software Engineering Conference*, 1995.
[6] Deiters, C., Herrmann, C., Hildebrandt, R., Knauss, E., Kuhrmann, M., Rausch, A., Rumpe, B., and Schneider, K., "GloSE-Lab: Teaching Global Software Engineering," in *Proceedings of 6th IEEE International Conference on Global Software Engineering*. IEEE, 2011.
[7] Keenan, E., Steele, A., and Jia, X., "Simulating Global Software Development in a Course Environment," in *International Conference on Global Software Engineering (ICGSE)*. IEEE, 2010.
[8] Richardson, I., Milewski, A. E., and Mullick, N., "Distributed Development — an Education Perspective on the Global Studio Project," in *International Conference on Software Engineering (ICSE)*. ACM, 2006.
[9] Huang, L., Dai, L., Guo, B., and Lei, G., "Project-Driven Teaching Model for Software Project Management Course," in *International Conference on Computer Science and Software Engineering*. IEEE, 2008.
[10] Dahiya, D., "Teaching Software Engineering: A Practical Approach," *ACM SIGSOFT Software Engineering Notes*, vol. 35, no. 2, 2010.
[11] Bavota, G., De Lucia, A., Fasano, F., Oliveto, R., and Zottoli, C., "Teaching Software Enginerring and Software Project Management: An Integrated and Practical Approach," in *International Conference on Software Engineering (ICSE)*. IEEE, 2012.
[12] Pádua, W., "Measuring complexity, effectiveness and efficiency in software course projects," in *International Conference on Software Engineering (ICSE)*. ACM, 2010.
[13] Ocampo, A. and Münch, J., "Rationale modeling for software process evolution," *International Journal on Software Process: Improvement and Practice*, vol. 14, no. 2, 2009.
[14] Kuhrmann, M., "A Practical Approach to align Research with Master's Level Courses," in *15th International Conference on Computational Science and Engineering*. IEEE, (to appear) 2012.
[15] Dukovska-Popovska, I, Hove-Madsen, V., and Nielsen, K. B., "Teaching lean thinking through game: Some challenges," in *36th European Society for Engineering Education (SEFI) on Quality Assessment, Employability & Innovation*, 2008.
[16] Münch, J., Pfahl, D., and Rus, I., "Virtual software engineering laboratories in support of trade-off analyses," *International Software Quality Journal*, vol. 13, no. 4, 2005.
[17] Joint Technical Committee ISO/IEC JTC 1, Subcommittee SC 7, "Systems and software engineering – Software life cycle processes," International Organization for Standardization, Tech. Rep. ISO/IEC 12207:2008, 2008.
[18] ISO TC22/SC3/WG16, "Road vehicles – Functional safety," International Organization for Standardization, Tech. Rep. ISO 26262:2011, 2008.
[19] IEC, "Medical device software – Software life cycle processes," International Electrotechnical Commission, Tech. Rep. IEC 62304:2006, 2006.
[20] Mendez Fernandez, D., Penzenstadler, B., and Kuhrmann, M., "Pattern-Based Guideline to Empirically Analyse Software Development Processes," in *16th International Conference on Evaluation & Assessment in Software Engineering*, 2012.
[21] Eclipse Foundation, "Eclipse Process Framework (EPF)," Online, http://www.eclipse.org/epf, 2010.
[22] Armbrust, O., Ebell, J., Hammerschall, U., Münch, J., and Thoma, D., "Experiences and results from tailoring and deploying a large process standard in a company," *Software Process: Improvement and Practice*, vol. 13, no. 4, 2008.

[23] K. Schwaber, *Agile Project Management with Scrum*. Microsoft Press, 2004.
[24] K. Beck, *Extreme Programming*. Addison-Wesley, 2003.
[25] P. Kruchten, *The Rational Unified Process: An Introduction*, 3rd ed. Addison-Wesley Longman, 2003.
[26] Federal Ministry of the Interior, "V-Modell XT Online Portal," Online, 2010. [Online]. Available: http://www.v-modell-xt.de/
[27] S. Brinkkemper, M. Saeki, and F. Harmsen, "Meta-Modelling Based Assembly Techniques for Situational Method Engineering," *Information Systems*, vol. 24, no. 3, 1999.
[28] R. Kneuper, *CMMI: Improving Software and Systems Development Processes Using Capability Maturity Model Integration (CMMI-Dev)*, 1st ed. Rocky Nook, 2008, no. ISBN: 978-3898643733.
[29] van Loon, H., *Process Assessment and ISO/IEC 15504: A Reference Book*. Springer, 2007.
[30] *ISO 9000:2005. Quality Management Systems – Fundamentals and Vocabulary*. International Organization for Standadization, 2005.
[31] Office of Government Commerce, *ITIL Lifecycle Suite 2011*. The Stationery Office Ltd., 2011.
[32] D. Rombach, "Integrated Software Process and Product Lines," in *International Software Process Workshop (SPW)*, ser. Lecture Notes in Computer Science. Springer, 2005.
[33] Deming, W. E., *Out of the Crisis*, 2nd ed. MIT Press, 2000.
[34] SCAMPI Upgrade Team, "Standard CMMI Appraisal Method for Process Improvement (SCAMPI) A, Version 1.3: Method Definition Document," Software Engineering Institute, Carnegie Mellon University, Tech. Rep. CMU/SEI-2011-HB-001, 2011.
[35] Joint Technical Committee ISO/IEC JTC 1, Subcommittee SC 7, "Software engineering – metamodel for development methodologies," International Organization for Standardization, Tech. Rep. ISO/IEC 24744:2007, 2007.
[36] OMG, "Software & Systems Process Engineering Metamodel Specification (SPEM) Version 2.0," Object Management Group, Tech. Rep., 2008.
[37] W. S. Humphrey, "The software process: Global goals," in *International Software Process Workshop (SPW)*, ser. Lecture Notes in Computer Science. Springer, 2005.
[38] A. Goodman, *Defining and Deploying Software Processes*. Auerbach Publishers Inc., 2005.